\documentclass[twocolumn]{aastex63}
\usepackage{graphicx}
\usepackage{amsmath}
\usepackage{amssymb}
\usepackage{float}
\usepackage{subfigure}
\usepackage{lipsum}
\usepackage{gensymb}
\usepackage{mhchem}
\usepackage{graphicx}
\usepackage{mathtools}
\usepackage{scalerel}
\usepackage{times}
\usepackage{verbatim}
\usepackage{tablefootnote}
\usepackage{savesym}
\savesymbol{tablenum}
\usepackage{longtable}
\usepackage{siunitx}
\restoresymbol{SIX}{tablenum}
\usepackage{multirow}
\usepackage{url}
\usepackage{hhline}
\newif\ifAMStwofonts

\newcommand\nustar{{\it NuSTAR}}
\newcommand\rosat{{\it ROSAT}}

\newcommand\suzaku{{\it Suzaku}}

\newcommand\xrism{{\it XRISM}}

\newcommand\swift{{\it Swift}}
\newcommand\xmm{{\it XMM-Newton}}

\newcommand\cdof{{\rm C/{\rm dof}}}

\newcommand\dec{{\Delta {\rm C}}}
\newcommand\feka{{Fe~K$\alpha$}}
\newcommand\fek{{Fe~K}}
\newcommand\oiii{{[O~\textsc{iii}]}}
\newcommand\feii{{Fe~\textsc{ii}}}

\newcommand\neix{{Ne~\textsc{ix}}}
\newcommand\nh{{N_{\rm H}}}
\newcommand\cts{{\rm\thinspace count}}
\newcommand\cps{\hbox{$\cts\s^{-1}\,$}}
\newcommand\pscm{\hbox{$\cm^{-2}\,$}}

\newcommand\arcs{{\hbox{$^{\prime\prime}$}}}
\newcommand\A{{\rm\thinspace \AA}}
\newcommand\cm{{\rm\thinspace cm}}

\newcommand\kev{{\rm\thinspace keV}}

\newcommand\m{{\rm\thinspace m}}
\newcommand\pc{{\rm\thinspace pc}}
\newcommand\Msun{\hbox{$\rm\thinspace M_{\odot}$}}
\newcommand\ks{{\rm\thinspace ks}}
\newcommand\s{{\rm\thinspace s}}

\newcommand\ps{{\rm\thinspace s^{-1}}}
\received{xxx yyy zzz}
\revised{xxx yyy zzz}
\accepted{Aug 17 2020}
\submitjournal{ApJ}

\shorttitle{Uncovering the primary X-ray emission in Mrk 1239}

\begin{document}

\title{Uncovering the primary X-ray emission and possible starburst component in the polarized NLS1 Mrk 1239}

\correspondingauthor{Margaret Buhariwalla}
\email{mbuhariwalla@ap.smu.com}

\author{Margaret Z. Buhariwalla}
\affiliation{Department of Astronomy \& Physics, Saint Mary's University, 923 Robie Street, Halifax, Nova Scotia, B3H 3C3, Canada\\}

\author{Sophia G. H. Waddell}
\affiliation{Department of Astronomy \& Physics, Saint Mary's University, 923 Robie Street, Halifax, Nova Scotia, B3H 3C3, Canada\\}

\author{Luigi C. Gallo}
\affiliation{Department of Astronomy \& Physics, Saint Mary's University, 923 Robie Street, Halifax, Nova Scotia, B3H 3C3, Canada\\}

\author{Dirk  Grupe}
\affiliation{Dept. of Physics, Earth Science, and Space System Engineering, Morehead State University, 235 Martindale DR, Morehead, KY 40351, USA\\}

\author{S. Komossa}
\affiliation{Max-Planck-Institut f{\"u}r Radioastronomie, Auf dem H{\"u}gel 69, 53121, Bonn Germany\\}



\begin{abstract}
X-ray observations of the unique NLS1 galaxy Mrk 1239 spanning 18-years are presented. Data from \xmm, \suzaku, \swift\ and \nustar\ are combined to obtain a broad-band, multi-epoch view of the source. There is spectral variability in the $3-10\kev$ band over the 18-years. An analysis of the \nustar\ and \suzaku\ light curves also suggests rapid variability in the $3-10\kev$ band, which is consistent with the NLS1 definition of the source. However, no variability is seen below 3 keV on any timescale. Two distinct physical models are adopted to describe the data above and below $\sim3\kev$. The low energies are dominated by a hot, diffuse gas likely associated with a starburst component at large physical scales. The higher energy spectrum is dominated by emission from the central region. Ionised partial covering and relativistic blurred reflection are considered for the central region emission. In both cases, the underlying power law has a photon index of $\Gamma\sim2.3-2.4$.  A distant reflector, a neutral partial covering component with a covering fraction near $\sim1$, and contributions from starburst emission are always required. The blurred reflection model requires a reflection dominated spectrum, which may be at odds with the low emissivity index and radio properties of the source. By contrast, the two absorption components required in the ionised partial covering model may correspond to the two distinct regions of polarization observed in the optical. Regardless of the physical model, spectral changes between epochs are driven by the absorption components and on short time scales by intrinsic AGN variability.

\end{abstract}



\section{Introduction} \label{sec:intro}
\label{sect:intro}
Active Galactic Nuclei (AGN) are powered by supermassive black holes which are accreting material from their surroundings. AGN emit light across the entire  electromagnetic spectrum and are typically variable at all wavelengths. Studying the X-ray emission from these objects allows for the characterisation of the innermost regions of the system, where extreme relativistic effects occur.

AGN are typically classified based on their viewing angle, where Seyfert 1 (type-1) AGN provide a direct view of the central engine, and Seyfert 2 (type-2) AGN are viewed through the cold, obscuring torus (\citealt{Antonucci+1993}). Seyfert 1 galaxies can then be further classified based on their optical properties, in particular, the full-width-half-maximum (FWHM) of the H$\beta$ line. Narrow-line Seyfert 1 (NLS1) galaxies have FWHM less than $2000~\textrm{km}\ps$, while broad-line Seyfert 1 (BLS1) galaxies have FWHM greater than $2000~\textrm{km}\ps$ (\citealt{Osterbrock+1985,Goodrich+1989}). The narrower lines observed in NLS1s are typically explained by lower mass AGN which are accreting at a higher fraction of their Eddington limit (\citealt{Pounds+1995,Grupe+2004, Komossa+2008}).

The X-ray spectra of type-1 AGN  is dominated by a power law. The origin of the power law is the X-ray emitting corona, a source of hot electrons located at some height above the black hole. UV seed photons from the accretion disc are Compton up-scattered in the corona and re-emitted as X-rays in the form of a power law. Many spectra also show evidence of a prominent emission line at $\sim6.4\kev$. This is typically attributed to \feka\ emission from neutral iron, originating in the torus, a distant cloud of neutral gas and dust (e.g. \citealt{Nandra+2007}). Above $10\kev$, spectra show evidence for a Compton hump, peaking at $20-30\kev$. This feature is produced via Compton down-scattering of photons from the corona in an optically thick medium, such as the torus or the accretion disc.

Below $\sim2\kev$, many Seyfert 1 AGN show a strong soft excess of disputed origin. This feature has been shown to be particularly prominent in NLS1 galaxies (e.g. \citealt{Boller+1996}; \citealt{Grupe+1998}; \citealt{Puchnarewicz+1992}). One commonly adopted interpretation is the partial covering scenario, wherein the soft excess is produced via absorption of X-rays from the corona (e.g. \citealt{Tanaka+2004}). The absorber is typically located close to the corona, and a number of ionisation states, densities, and covering fractions of the absorbing material can be adopted to explain the observed emission. This interpretation has been used to model the observed spectra of numerous type-1 AGN (e.g. \citealt{Miyakawa+2012,Gallo+2015}).

An alternative interpretation is known as the blurred reflection scenario (e.g. \citealt{RossFabian+2005}). In this model, some fraction of photons emitted by the corona are incident upon the accretion disc. As the photons interact with the disc, they produce strong emission and absorption features, most of which are associated with iron and have energies below $\sim2\kev$. As the disc rotates, the material is subject to extreme general relativistic effects, and the features appear broadened. This produces a strong soft excess, as well as a broadened \feka\ line between $4-7\kev$. This model has been successfully applied to the spectra of numerous NLS1 galaxies (\citealt{Fabian+2004,Ponti+2010,Gallo+2019m335}).

When the continuum (power law) component is significantly suppressed, for whatever reason, the underlying components can often be distinguished.  In dim sources, distant reflection from the torus and even X-ray emission from star formation in the galaxy can also contribute to the soft excess (e.g. \citealt{Franceschini+2003,Gallo+2019m335,Parker+2019}) . While the partial covering and blurred reflections typically produce smoother soft excesses, the torus and star formation regions lie far away from the central engine and are not subject to extreme relativistic effects. The emission and absorption features therefore appear narrow.

Mrk 1239 (RX J0952.3-0136) is typically classified  as a NLS1 galaxy and is found at a redshift of $z=0.01993$ (\citealt{Beers+1995}). The source has been studied at many wavelengths and numerous interesting properties have been revealed. The mass of the  supermassive black hole at the centre of Mrk 1239 has been reported as $2.4\times10^6 \Msun$ (\citealt{Marin+2016}).  \cite{VC+2001} measure a FWHM of $1075~\textrm{km}\ps$ for the H$\beta$ line, moderate \oiii\ emission strength (\oiii/H$\beta$ = 1.29), and weak \feii\ emission ([\feii]/H$\beta$ = 0.63).  Additionally, the optical spectrum shows evidence for polarisation on the order of $\sim3-4$\%, and has one of the highest degrees of polarisation reported in \cite{Martin+1983}. \cite{Goodrich+1989} show that the Balmer lines and forbidden optical lines show different degrees of polarisation and suggest that these features have distinct physical origins, polarised due to dust reflection and transmission.

In the radio regime, Mrk 1239 has been classified as both radio-quiet (\citealt{Do+2015}), or borderline radio-loud (\citealt{Berton+2018}). \cite{Do+2015} show that the radio emission cannot solely be attributed to starburst activity and must also comprise AGN jet activity. They classify this source as a Fanaroff-Riley Type I candidate (\citealt{Fanaroff+1974}), meaning that the radio luminosity decreases with increasing distance from the centre of the galaxy. \cite{Do+2015} also give clear evidence for kilo-parsec scale non-thermal radio emission attributed to AGN jets, however, most of the radio power is centred in the inner $100\pc$.

Based on near-infrared observations, Mrk 1239 has some evidence for star forming regions based on signatures from polycyclic aromatic hydrocarbon (PAH). \cite{Ruschel_Dutr} and \cite{Jensen+2017} both report signatures of PAH at 11.3$\mu$m, although both suggest that these features do not lie in the inner nucleus, but rather a few hundred pc from the centre. \cite{Rod+2003} place only an upper limit on a $3.3\mu$m PAH detection,  also state  that starburst activity may be occurring a few hundred pc from the nucleus. An estimated star formation rate (SFR) of less  than $7.5\Msun \rm yr^{-1}$ (\citealt{Ruschel_Dutr}) has been measured. \cite{Sani+2010} find that Mrk 1239 exhibits weaker star formation relative to AGN emission than the average NLS1. As well, \cite{Rod+2006} reported a remarkable NIR bump, they interpret it as a massive reservoir of dusty  gas between NLR and BLR. This could explain strong continuum absorption.

In the X-ray, observations with \rosat\ and \xmm\ have previously been analysed. \cite{Rush+1996} report a soft X-ray slope of $\Gamma\simeq3$ using \rosat\, and find absorption higher than the Galactic $\nh$ value by a factor of $\sim1.5$. Mrk 1239 is also included in the \rosat\ sample analysed by \cite{Boller+1996}, where a steep soft X-ray slope of $3.9$ and a high column density of $8.3\times10^{20}\pscm$ is reported.
\cite{Grupe+2004} report on a $10\ks$ \xmm\ observation of Mrk 1239, using data from the EPIC-pn and MOS detectors. They find that the spectral shape can successfully be reproduced using a power law which is almost entirely absorbed by two distinct absorbers, akin to the two polarisation regions reported in \cite{Goodrich+1989}. They also report a strong feature around $0.9\kev$ found in all three detectors, which they attribute to a strong \neix\ line due to a super-solar Ne/O ratio (\citealt{Grupe+2004}).

This work presents the spectral and timing analysis of all available X-ray data from \xmm, \suzaku, \nustar\ and \swift, spanning 18 years between 2001 and 2019, and seeks to explain the unique X-ray properties of the source. In Section \ref{sect:data}, the observations and data reduction techniques are summarised. Section \ref{sec:VAR} examines the variability of the source across both long (years) and sort (hours) timescales  and in Section \ref{sec:MO} the spectra are analyzed. A discussion of the results is given in Section \ref{sec:Disc}, and conclusions are drawn in Section \ref{sect:conclusion}.

\section{Observations and Data Reduction}
\label{sect:data}
Mrk 1239 was observed with \xmm, \suzaku, and \nustar/\swift\ at three different epochs over 18 years. The data analysed here are listed in  Table \ref{tab:obs}. This section describes the observations and data reduction.
\begin{table*}
	\centering
	\begin{tabular}{c c c c c c c c}
		\hline
		(1) & (2) & (3) & (4) & (5) & (6) & (7) & (8)\\
		Observatory & Observation ID & Name & Start Date   & Duration & Exposure & Counts  	  & Energy Range\\
		& & & (yyyy-mm-dd) & [s]      & [s]      &       	  & [keV]\\
		\hline
		\xmm\ MOS 1+2	  		&0065790101	& MOS    &2001-11-12    &  	9959    & 9371/9371   &  1052	  & 0.3-8.0  \\
		\hline
		\suzaku\ XIS 0+3 	&702031010 	& XIS 	 & 2007-05-06   &  	126256	& 63128  &  6555	  & 0.7-1.5, 2.5-10  \\
		\suzaku\ PIN    	&   	   	& PIN	 &		   		&  	535904	& 		 &  14654	  & 12-20 \\
		\hline
		\nustar\ FPMA/FPMB	&60360006002& FPMA/FPMB&2019-06-17    & 	38053	& 21093	 & 2039/1954  & 3-30 \\
		\hline 
		\swift\ XRT 		&00081986001& XRT	 &2019-06-17    & 	18624	& 6216   & 	183	  	  &	0.5-7 \\
	\end{tabular}
	\caption{Observations log for Mrk 1239. The observations and  instruments used for analysis are listed in column (1). The observation ID and labels used in this work are given in columns (2) and (3), respectively. The start date of each observation is given in column (4). The duration of each observation, total exposure time and total counts for each observation are given in columns (5), (6), and  (7), respectively. The energy each observation were fit over is given in column (8). For \suzaku\, the combined counts from XIS0 and XIS3 are given (column 7). Similarly for \xmm\ the combined counts from MOS1 and MOS2 is reported.}
	\label{tab:obs}
\end{table*}

\subsection{\xmm}
\label{sect:xmm}
Mrk 1239 was observed with \xmm\ (\citealt{Jansen+2001}) for $\sim10\ks$ in late 2001. The source appears in the field of view of the target source, RXJ 095208.7-014818, and is therefore substantially off axis. 

The \xmm\ Observation Data Files (ODF) were processed to produce a calibrated event list using the \xmm\ Science Analysis System, {\sc sas v17.0.0}. Examination of the background showed significant  flaring in the EPIC-pn detector. A good time interval (GTI) was created and applied. Background flaring was not significant in the EPIC-MOS1 and MOS2 detectors.

For each detector, source photons were extracted from a circular region with a $35\arcs$ radius centred on Mrk 1239, and background photons were extracted from an off-source circular region with a $50\arcs$ radius on the same CCD. For the pn detector, single and double events were selected, while single to quadruple events were selected for the MOS detectors. The {\sc sas} tasks {\sc rmfgen} and {\sc arfgen} were used to generate response files. \xmm\ light curves were not  examined due to the short length  of the observation (10 ks).

The source and background spectra were  binned with a minimum of 10 counts per bin. The final pn, MOS1 and MOS2 spectra were checked for consistency, and all spectra were found to be comparable within known uncertainties.\footnote{\url{https://xmmweb.esac.esa.int/docs/documents/CAL-TN-0018.pdf}}
The pn spectrum had low counts due to background flare filtering and the source was located at the very edge of the detector. For this reason, the combined MOS data are used for further analysis. The ungrouped MOS1 and MOS2 source and background spectra were merged using {\sc addspec}, and the corresponding response files merged using {\sc addrmf} and {\sc addarf}. The combined source and background spectra were then binned with a minimum of 10 counts per bin. This combined MOS spectrum had a higher signal and lower background then the pn detector, allowing for improved spectral modelling. Data above $8\kev$ are background dominated so only the $0.3-8\kev$ range is used for spectral modelling.
\subsection{\suzaku}
\label{sect:suz}
Mrk 1239 was the target of a $\sim126\ks$ \suzaku\ (\citealt{Mitsuda+2007}) observation in May 2007. The data were taken in XIS nominal mode. Extraction of spectra and light curves was performed with {\sc xselect v2.4g} using cleaned event files from the front illuminated (FI; XIS0 and XIS3) and back illuminated (BI; XIS1) CCDs. 

For each instrument, source photons were extracted using a $240\arcs$ region centred around the source, while background photons were extracted from a $180\arcs$ off-source region, while avoiding the calibration regions in the corners of the CCDs.  Response matrices for each detector were generated using the tasks {\sc xisrmfgen} and {\sc xissimarfgen}. Source and background spectra for each detector were then  binned using the optimal binning routine  in {\sc ftgrouppha} (\citealt{Kaastra_2016}). The XIS spectra were checked for consistency and found to be comparable with one another. The  XIS0 and XIS3 source and background spectra were then merged using {\sc addascaspec}. For simplicity, only the merged FI spectra are presented for the remainder of the analysis.
The XIS spectrum is modelled between 0.7- 10\kev, while excluding $1.5-2.5 \kev$ due to calibration uncertainties.

Cleaned event files from the HXD-PIN detector were processed using the tool {\sc hxdpinxbpi}, resulting in a $54\ks$ exposure. Both the non-X-ray background (NXB) and the cosmic X-ray background (CXB) are used to determine the background level. The source is detected at $\sim 4.5$ per cent between 12-20 \kev, which is considered marginal (\citealt{Fukazawa+2009}).

\subsection{\nustar\ and \swift}
\label{sect:nuStar}
Mrk 1239 is part the \nustar\ (\citealt{nustar+2013}) Extragalactic Legacy Survey, specifically the  \nustar\ Local AGN NH Distribution Survey (NuLANDS).  NuLANDS is designed to look at heavily obscured AGN in the local universe (\citealt{Boorman+2018}).  There are 30 AGN in the sample and observations  were completed in 2019.

Mrk 1239 was observed in June 2019 with a simultaneous \swift\ observation shortly after \nustar\ began. FPMA and FPMB data were extracted from source region of 75". A background was selected from the same chip with a region of 115". The data were processed with {\sc caldb} index version 20181030. The joint \swift\ XIS spectrum was obtained from  the \swift-XRT data product generator (\citealt{Evans+2009})\footnote{\url{www.swift.ac.uk/user_objects/}}. 

The \nustar\  data were optimally binned using {\sc ftgrouppha}. The \swift\ spectra  was binned to have a minimum of 10 counts per bin using {\sc grppha}.



\section{Characterizing the Variability}
\label{sec:VAR}
\subsection{Long term Variability}
\begin{figure}
	\includegraphics[width=\columnwidth]{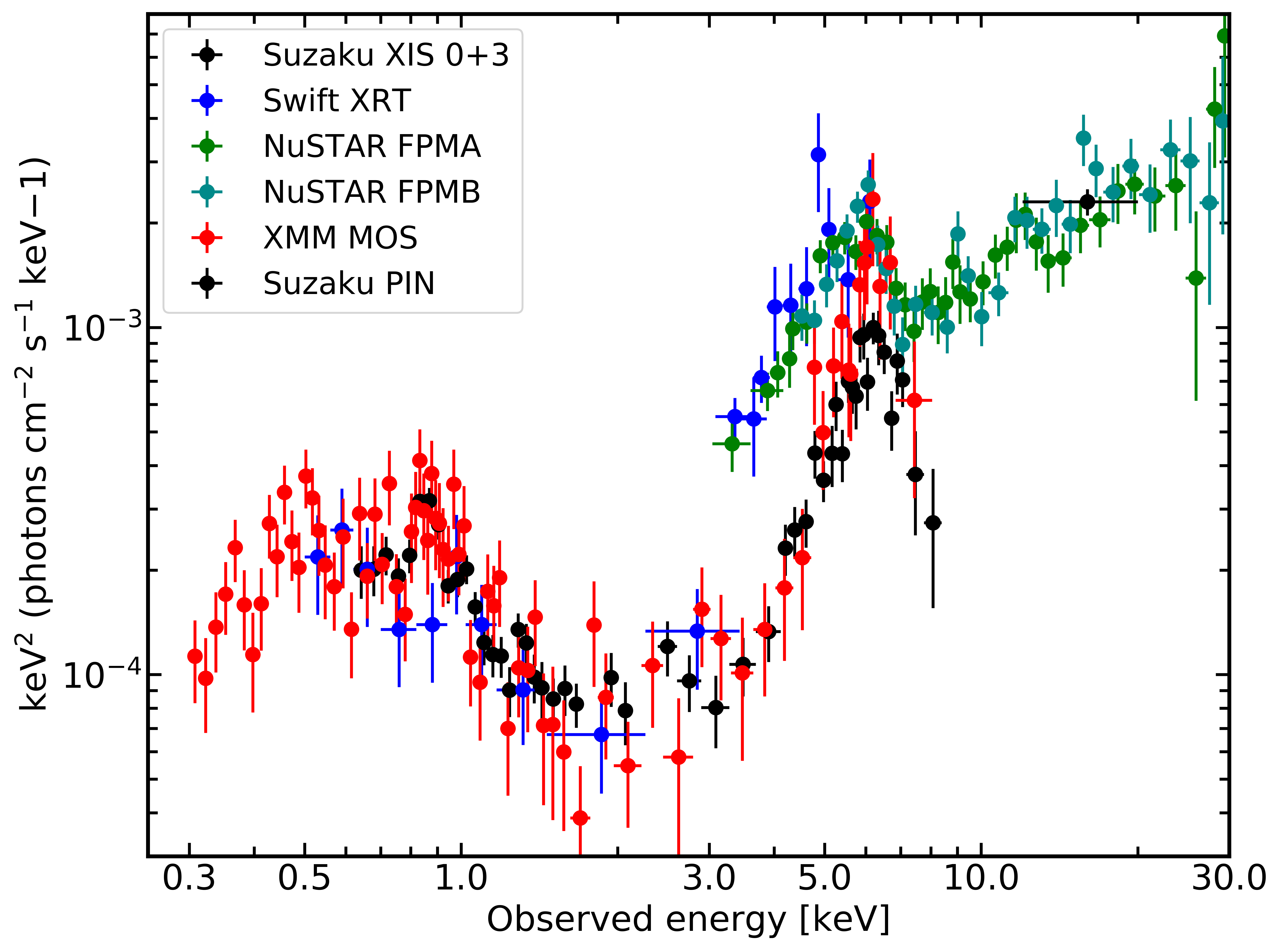}
	\caption{The unfolded spectra of Mrk 1239, with observations spanning 18 years. Data from \xmm\ are shown in red, \suzaku\ are shown in black, \nustar\ are shown in green and aqua, and \swift\ are shown in blue. The data are compared to a flat power law with $\Gamma = 0$. The intrinsic spectrum is hard, and spectral changes are apparent in the $3-10\kev$ band. }
	\label{fig:eeuf}
\end{figure}
The unfolded spectra of Mrk 1239 compared to a flat ($\Gamma = 0$) power law are plotted in Figure~\ref{fig:eeuf}. Plotting the unfolded spectra in this way allows us to directly compare data from different instruments, highlighting any spectral differences at each of the three epochs.

The soft band ($0.3-3\kev$) is remarkably similar between epochs, displaying the same flux and curved spectral shape in the XRT, XIS and MOS spectra. The large hump like feature appears to be a soft excess. This feature has been consistently observed in this source outside the observations included in this paper. The source was also observed using \textit{ROSAT} (\citealt{rosat}) in 1992. \cite{Rush+1996} fit the source with an absorbed power law and found that was an acceptable fit. However the fit was improved using an additional component, either thermal emission (black body or Raymond-Smith thermal plasma) or an emission line at $\sim0.7\kev$. The spectra lacked the data quality to distinguish between any of the models. An emission line-like feature at $\sim0.9\kev$ was also reported by \cite{Grupe+2004} in the EPIC-pn spectra. This feature is  prevalent in all the data. 

The most significant variability appears between $3-10\kev$. The flux and spectral shape of the source is different in 2019 compared to the earlier XIS and MOS observations. In particular, the \nustar\ and \swift\ data are a factor of $\sim2-3$ brighter than the XIS and MOS data in this energy range. The MOS and XIS data are remarkably similar across the entire energy range. There is a slight discrepancy around $\sim6-7\kev$, where the MOS data appear slightly brighter; however, the spectra remain consistent within error.

The source is significantly detected in \nustar\ up to $\sim30\kev$. The source is only marginally detected in the  PIN, but appears to be comparable in brightness at the two epochs (2007 and 2019). We cannot comment on any change in shape between epochs because the PIN data are limited.

\subsection{Rapid variability}
\label{sect:var}
\begin{figure*}
	\begin{minipage}{0.48\textwidth}
		\includegraphics[width=\columnwidth]{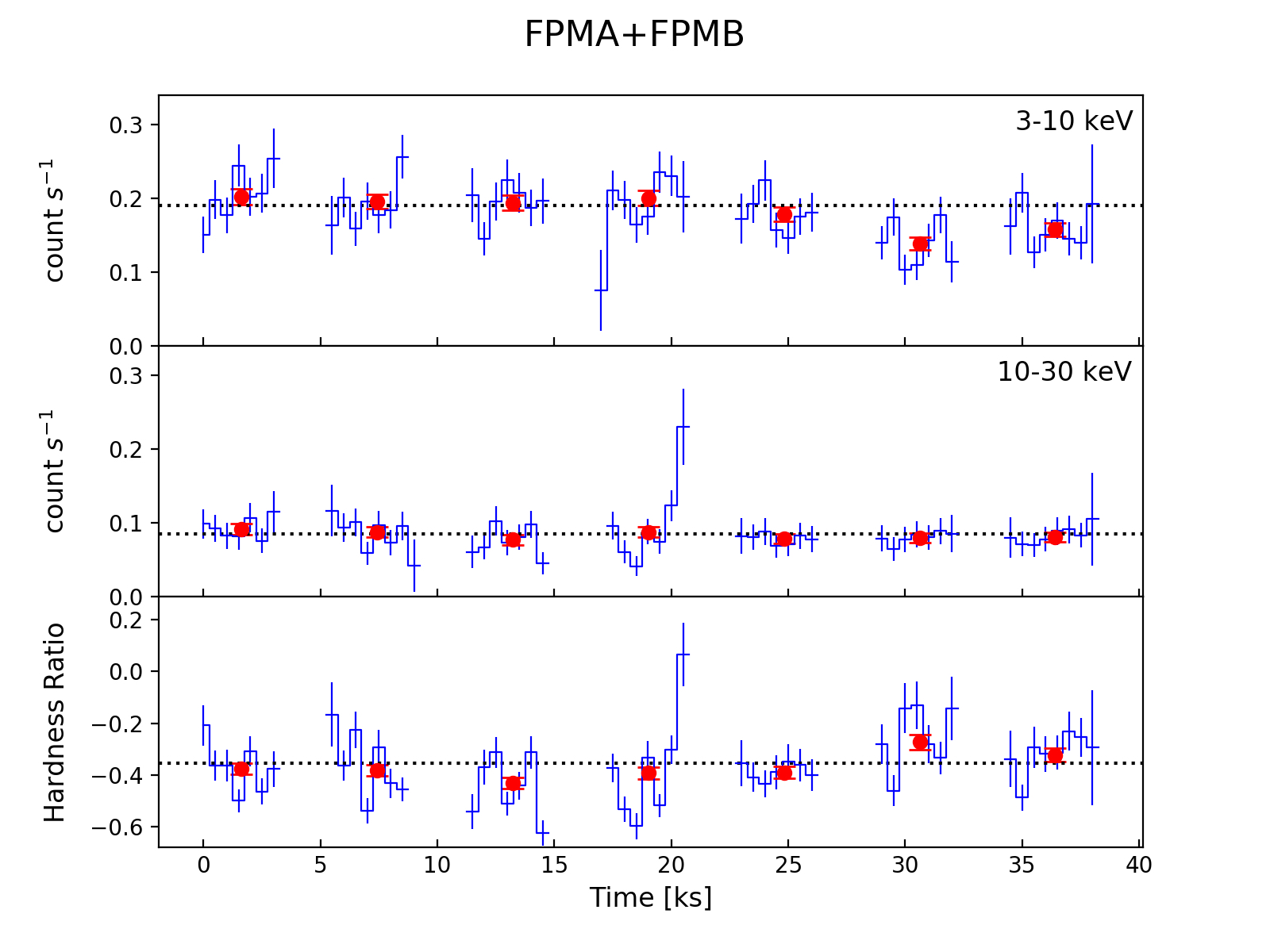}
	\end{minipage}
	\begin{minipage}{0.48\textwidth}
		\includegraphics[width=\columnwidth]{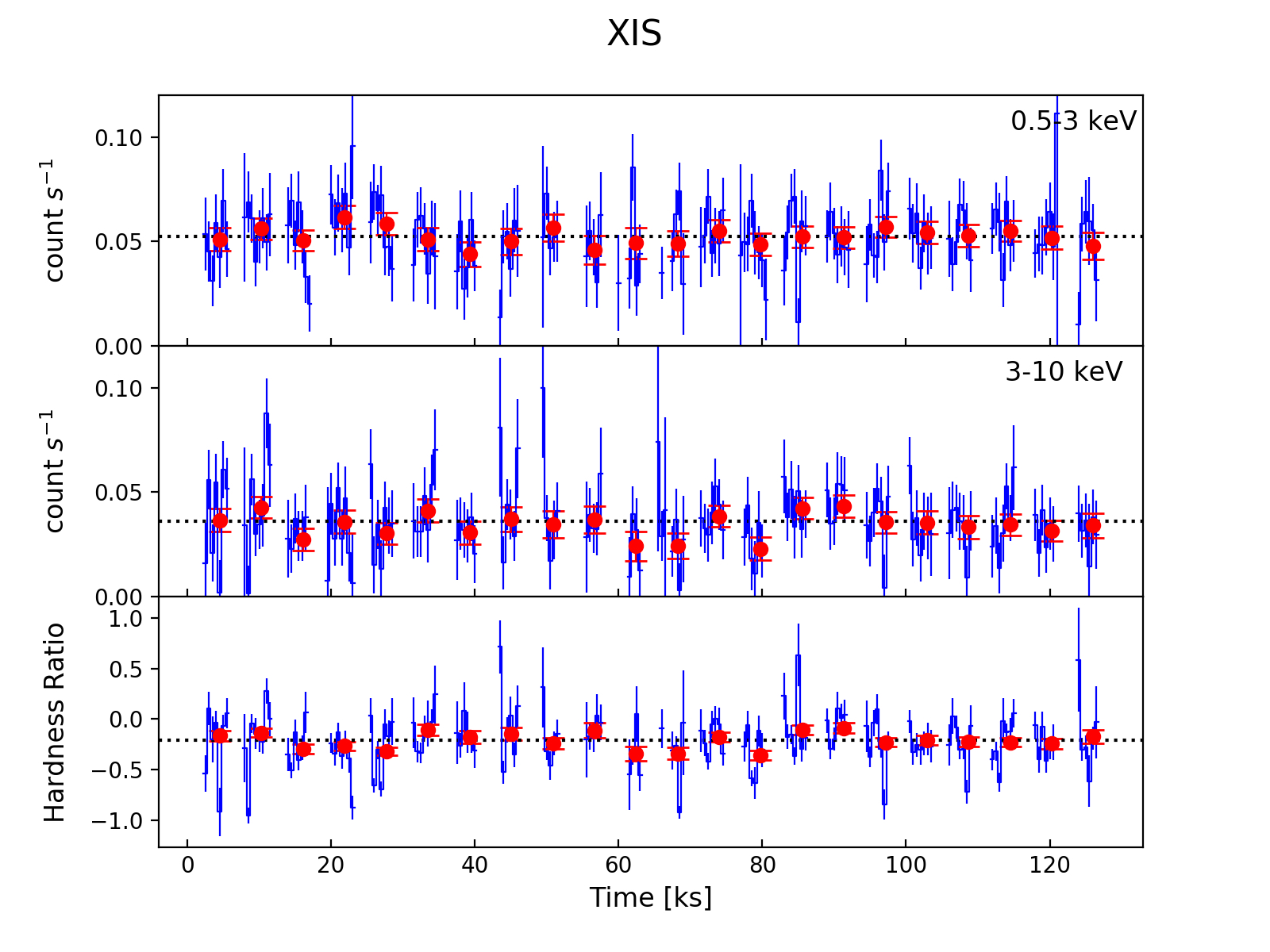}
	\end{minipage}
	\caption{\textit{Left:} \nustar\ light curve with FPMA and FPMB combined. Top panel is the 3-10 \kev\ band, middle is the 10-30 \kev\ band and bottom panel is the hardness ratio calculated by: $HR=(H-S)/(H+S)$. \textit{Right:} \suzaku\ light curve. Top panel is the 0.5-3 \kev\ band, middle is the 3-10 \kev\ band and bottom panel is the hardness ratio. In all panels the dashed line is the average count rate or HR. }
	\label{fig:refcorr}
\end{figure*}

Having established that there is some long-term variability in the $3-10\kev$ band, we examine light curves for rapid variability over the course of the \nustar\ and \suzaku\ observations. The FPMA and FPMB lightcurves were merged to improve signal-to-noise, as were the XIS0 and XIS3 lightcurves. Variability was examined using $500\s$ and $5780\s$ bins, where $5780\s$ corresponds approximately to the orbital period of the satellites. The lightcurves were binned using {\sc lcurve}. Given the modest data quality, all lightcurves and hardness ratios are compared to their mean value to examine variability. The fit quality is given by a reduced $\chi^2$ test. In Figure \ref{fig:refcorr} and Table \ref{tab:chi}, the light curves and fits are presented.

\def\arraystretch{1.5}
\begin{table}
	\centering
	\begin{tabular}{c c c c }
		\hline
		(1)		&(2)			& \multicolumn{2}{c}{(3)}\\
		Instrument	&Energy band 	&\multicolumn{2}{c}{$\chi^2 / dof$}  \\
		& 				& $500\s$		& $5780\s$	\\
		\hline
		FPMA+FPMB& $3-10\kev$		& 101/50	&41/6	\\
		& $10-30\kev$	& 53/50		& 4/6	\\
		& ${\rm HR_{FPM}}$	& 198/49 	&25/6 	\\
		\hline
		XIS & $0.5-3\kev$		&  144/145 	&11/21	\\
		& $3-10\kev$		&  139/142 	&23/21	\\
		& ${\rm HR_{XIS}}$	&  658/139	&52/21	\\
		\hline
	\end{tabular}
	\caption{The fit statistic for each lightcurve and HR  compared to their mean value. Column (1) indicates which instrument the light curve was produced by. Column (2) states the energy bin of the lightcurve. Column (3) gives the fit quality when the mean is fit for each light curve. }
	\label{tab:chi}
\end{table}

The XIS lightcurve for the soft band ($0.5-3\kev$) has a mean count rate of $\sim 0.05\cps$. The $\chi^2 / dof$ for the light curves is 144/145 using the $500\s$ bins and 11/21 using the $5780\s$ bins. This indicates that the source is not variable below $3\kev$ on short time scales.

The hard band ($10-30\kev$) was only observed with \nustar, and has a mean count rate of $\sim 0.1\cps$. The $\chi^2 / dof$ of the $500\s$ and $5780\s$ bin lightcurves are 53/50 and 4/6, respectively. The slight difference in the $\chi^2$ values may be the result of a flare in the $500\s$ bin lightcurve at approximately $20\ks$.  On short time scales, the hard band remains constant in flux within uncertainties

The intermediate band ($3-10 \kev$) is the only band that has overlap between \nustar\ and XIS. In the XIS lightcurve the mean count rate is $\sim0.04\cps$.  The AGN is more significantly detected  in \nustar\ and the mean count rate is $\sim 0.2\cps$. Upon visual inspection the intermediate band appears to have more variability compared to the soft and hard bands.  This is confirmed in the constant fit test.  The \nustar\ $500\s$ and orbital binned light curves are inconsistent with constants and have  $\chi^2 / dof$ = 101/50 and 41/6, respectively. The XIS data shows marginal variability (23/21) in the orbital binned lightcurve in this band and none in the $500\s$ bin lightcurve. However due to the low count rate it is difficult to rule out rapid variability in the XIS lightcurves.

The hardness ratio (HR) for the XIS lightcurves,  ${\rm HR_{XIS}}$, has a mean value of $\sim-0.2$, indicating  that the soft count rate is  higher than the intermediate count rate. The \nustar\ ${\rm HR_{FPM}}$ has a mean value of $\sim-0.4$, indicates the intermediate count rate is higher than the hard count rate.  All HR curves are inconsistent with a constant fit ($\chi^2/dof > 1$) indicating a significant amount of spectral variability.  This spectral variability is driven by fluctuations in the intermediate, $3-10\kev$, band. This is the same result found on long time scales (Figure~\ref{fig:eeuf}). On both time scales there is very little change in the soft band, which remained constant in shape and flux. The intermediate band changes in shape and flux on long time scales, as well it changes in flux on rapid time scales. 

The excess variance ($\sigma^{2}_{\rm rms}$) (e.g. \citealt{Ponti+2012}) is calculated for the  \nustar\ $3-10 \kev$ light curve to further examine the rapid variability.  The light curve is binned in $200\s$ and subdivided in $10\ks$ segments.  The excess variance is calculated in each segment and then averaged over all of the segments.  The measured average value for Mrk 1239 during the \nustar\ observation is $\sigma^{2}_{\rm rms}=0.04\pm0.01$. The value is comparable to that measured in other NLS1 galaxies (\citealt{Ponti+2012}).

Seyfert 1 AGN can display substantial variability at lower energies, and NLS1s in particular are known to display extreme soft variability on both short and long timescales (e.g. \citealt{Boller+1996}; \citealt{Leighly1999}; \citealt{Grupe+2001}; \citealt{Nikolajuk+2009}; \citealt{Grupe+2010}; \citealt{komossa+2016}; \citealt{Bonson+2018}; \citealt{Gallo+2018}).   The differences in the rapid and long term spectral variability may be pointing to different origins for the soft ($<3\kev$) and intermediate ($3-10\kev$) emission.  This suggests that the primary source of the soft excess is highly atypical compared to other NLS1s. However, the variability above $3\kev$ seen on both short and long timescales is more typical of what is seen in other sources, implying a distinct origin from the soft excess. More discussion for this will be given in Section~\ref{sec:Disc}.

\section{Spectral Modelling}
\label{sec:MO}
All data were background modelled in {\sc XSPEC}, except for the PIN data which was background subtracted. We used C-statistics (\citealt{Cash1979}) to evaluate the fit quality throughout. Errors were calculated at the 90 per cent confidence level using the {\sc XSPEC} error comand. The \nustar\ and \swift\ spectra were treated as one epoch for all spectral modelling. All model parameters were linked between them. The same was done for the \suzaku\ XIS and PIN data. A cross-calibration constant was applied to the FPMB detector in \nustar, the XRT data from \swift\ and the PIN data from \suzaku . The constant was free for FPMB and XRT and monitored to ensure it remained within acceptable ranges, as prescribed by \nustar\ FAQ\footnote{\url{https://heasarc.gsfc.nasa.gov/docs/nustar/nustar_faq.html}}. The constant applied to the PIN data was frozen at 1.18\footnote{\url{https://heasarc.gsfc.nasa.gov/docs/suzaku/analysis/abc/node8.html} (See Sec. 5.7.2)}. A Galactic column density of $\nh =\num{4.43 e+20}$ (\citealt{Willingale+2013}) was applied to all models and frozen throughout spectral fitting. \cite{Wilms+2000} abundances were used throughout spectral fitting.
\begin{figure*}
	\includegraphics[width=2\columnwidth]{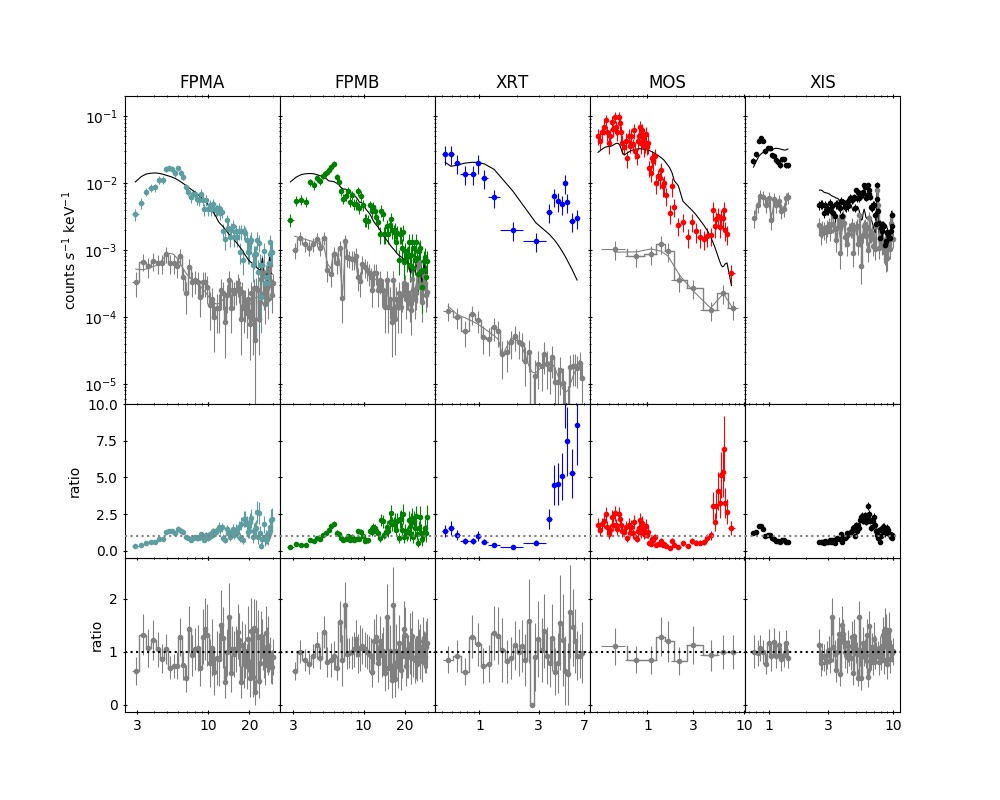}
	\caption{\textit{Top row:} The folded source spectra for each instrument with the background shown in grey. The model applied is a  power law with Galactic absorption, with $\Gamma=2$ and the normalization free for all instruments.  This model was applied to emphasize the need for more complex models than a simple power law.\textit{ Middle Row:} The ratio of the data with the simple power law model. Comparison to the dotted line at ${\rm ratio=1}$ demonstrates the curvature in the data. \textit{Bottom Row:} The ratio of the background to the background model, with ${\rm ratio=1}$. }
	\label{fig:apend_July_9}
\end{figure*}

Figure \ref{fig:apend_July_9} shows each spectra separately with its background model, and ratio for an absorbed power law with $\Gamma = 2$ to represent a typical NLS1. As we can see in Figure \ref{fig:apend_July_9},  all spectra exhibit curvature and none are fit well by the power law. All five spectra show evidence for excess emission between $5-8\kev$. The MOS spectrum shows a soft excess below 1 \kev\ and all spectra are over estimated by the model in the $2-5\kev$ range.

\subsection{\nustar /\swift}
\label{sect:NuSwiftmodel}
We begin by examining only the \nustar\ and \swift\ spectra, and only consider data above $2\kev$. This is the first time this data set has been examined in detail and it provides the best view of the AGN above $10\kev$. 
The examination of the unfolded spectrum reveals many key characteristics; an underlying power law component, extreme spectral curvature around $3-8\kev$, and a potential narrow feature at $6.4\kev$. To model these features, we apply a power law plus a  cold (log$\xi=0$; where the ionisation parameter $\xi= L/nr^2$ and $L$ is the incident luminosity, $n$ is the column density of the cloud, and $r$ is the distance from the cloud to the illumination source), distant reflector representing reflection off the torus. To reproduce the observed spectral curvature we modify the power law with a single region of neutral absorption. The {\sc XSPEC} model would appear as: {\sc constant$\times$tbabs$\times$(xillver + zpcfabs$\times$po)}. The constant is the cross calibration constant between FPMA and FPMB and \swift. This model was found to provide a suitable fit to the \nustar~/~\swift\ data in the $2-30\kev$ band, and gives a fit statistic of $\cdof= 356.46/314$. 

We extended the model and data down to $0.5\kev$ to see if this model is also capable of explaining the spectrum at low energies. This results in a much poorer fit ($\cdof = 449.86/338$) and a large excess below approximately $1\kev$. The addition of a second neutral partial covering component does not improve the fit quality,  however the addition of an ionized partial covering absorber ({\sc zxipcf}; \citealt{Reeves+2008}) significantly improves the fit quality  ($\cdof = 324.03/335$).  Despite the large change in the C-statistic, there remain significant residuals at approximately $1\kev$, as were noted by \cite{Grupe+2004}. Additionally, all model fits give very high values of the absorber covering fraction ($CF\sim0.95-1$). It therefore seems possible that the intrinsic emission from the AGN is largely obscured below $\sim3\kev$, and appears only at high energies.

Based on the work of \cite{Ruschel_Dutr} in the mid-infrared, we can infer the existence of star forming regions in Mrk 1239 (see Section \ref{sect:intro}).  Given the lack of evidence for variability at low energies, it is possible that the soft emission does not originate close to the central engine, but rather from star formation on extended scales. We therefore use {\sc mekal} (\citealt{Liedahl+1995}), collisionally ionised emission from hot diffuse gas, to model the SFR.  We specify that {\sc mekal} use abundances given by \cite{Wilms+2000}, with abundance frozen for all spectral fits. This improves the fit to $\cdof = 307.92/333$, and removes the residuals at $1\kev$. The measured temperature is ${\rm kT}\approx 1 \kev$. 

Alternatively, we examine if the curvature and hard X-ray emission could be attributed to blurred reflection (eg. \citealt{RossFabian+2005}) rather than the ionized partial covering. By removing the ionized absorber and replacing it with a blurred reflection ({\sc relxill}, \citealt{Garcia+2014}) one gets an acceptable fit with a simple blurred reflection model. A neutral partial covering component and {\sc mekal} are still needed. Furthermore, the use of \nustar\ and \swift\ spectra alone does not allow for the constraint of many of the blurred reflection parameters. This model  will be explored in more detail using additional data in Section~\ref{sec:refl}. 

\subsection{Multi-epoch Spectral Modelling}
In this section, we attempt to describe the behaviour of Mrk 1239 in a self consistent manner using the multi-epoch spectral data collected over 18 years. This has the advantage of giving us low energy sensitivity, provided by MOS and XIS, for the soft excess and the \feka\ region. \nustar\ provides us with good energy coverage up to $30\kev$, with simultaneous coverage between 0.5 and $7\kev$ from \swift. This allows us to simultaneously study the soft excess, \fek\ region and the broad band continuum. We motivate the models used based on spectral features and variability on multiple time scales.

Structure around the neutral \feka\ band seen in all spectra are suggestive of distant cold reflection, likely originating in the torus. For this we use {\sc xillver} with ${\rm log} \xi = 0$ to represent a cold reflector. 

Study of the long term and rapid variability shows that there is negligible variability below $3\kev$. This indicates that direct AGN continuum emission may be highly absorbed and thus not visible. There is structure seen in the MOS and XIS spectra that may be from distant optically-thin emission from star formation activity. We use {\sc mekal} for this feature. 

It also seems possible that the direct AGN component is revealed in the 3-10 \kev\ band, but completely obscured below $\sim3 \kev\ $ This indicates the presence of a partial covering absorber ({\sc zpcfabs}).  We expand upon Section~\ref{sect:NuSwiftmodel} by applying the ionized partial covering model and the blurred reflection model to all epochs. Some intrinsic properties of the system were linked, as they are not expected to change on timescales of years. This included the {\sc mekal} component (both normalization and kT), the distant cold reflector ({\sc xillver}) and some of the blurred reflection parameters discussed in detail in Section~\ref{sec:refl}.

Upon analysis of Figure \ref{fig:eeuf} and fitting both models it was found that the \suzaku\ spectra closely followed the MOS spectra. Linking all parameters between \suzaku\ and MOS did not significantly decrease the fit quality for either physical model, so they are linked throughout the remainder of the analysis. Allowing for a free constant between the \xmm\ and \suzaku\ models is not a significant improvement to the fit ($\dec=3$ for one additional free parameter). This leaves us with two epochs; a historic one containing \xmm\ MOS, \suzaku\ XIS and PIN data, and a  recent one containing the \nustar\ FPMA and FPMB and \swift\ XRT data.

\subsubsection{Partial Covering}
In a partial covering scenario a cloud of absorbing material with column density ($\nh$) is positioned in the line of sight of the AGN, obscuring a fraction of the direct emission. The model has some fraction of its source emission absorbed and the rest is considered direct emission  (e.g. \citealt{Holt+1980}; \citealt{Tanaka+2004}). The model includes three parameters: the column density ($\nh$); the covering fraction (CF), which is the fraction of  intrinsic emission that is absorbed; and the redshift ($z$), which remains fixed at $z=0.01993$. The direct emission can be either scattered off the clouds or let through patches in the absorbing material (\citealt{Tanaka+2004}). The patchy absorber is often called a ``leaky absorber" (\citealt{Tanaka+2003}) as the direct emission photons can ``leak" through the absorber. 

Partial covering is often employed to fit NLS1s (e.g. \cite{Boller_PC}; \citealt{Tanaka+2005}; \citealt{Turner+2007}; \citealt{Gallo+2015};  \citealt{Iso+2016}; \citealt{Grupe+2019}.  For Mrk~1239, \cite{Grupe+2004} use two absorbing components to model a leaky absorber with the EPIC-pn spectrum. A variation of partial covering is ionized partial covering, where the absorbing medium has some ionization. In this scenario the illumination source is the power law corona (e.g. \citealt{Reeves+2008}).
\begin{figure}
	\begin{minipage}{0.48\textwidth}
		\includegraphics[width=\columnwidth]{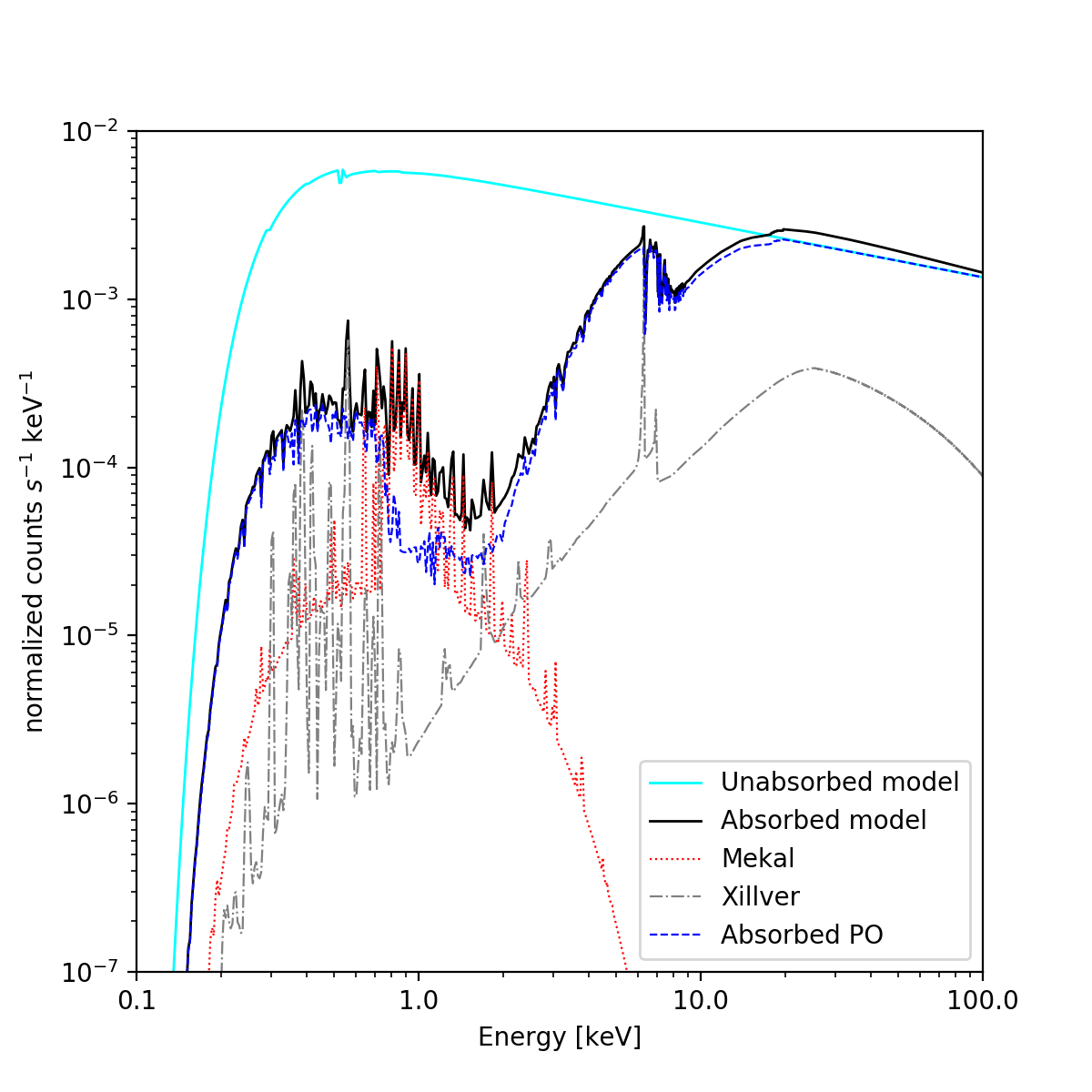}
	\end{minipage}
	\begin{minipage}{0.48\textwidth}
		\includegraphics[width=\columnwidth]{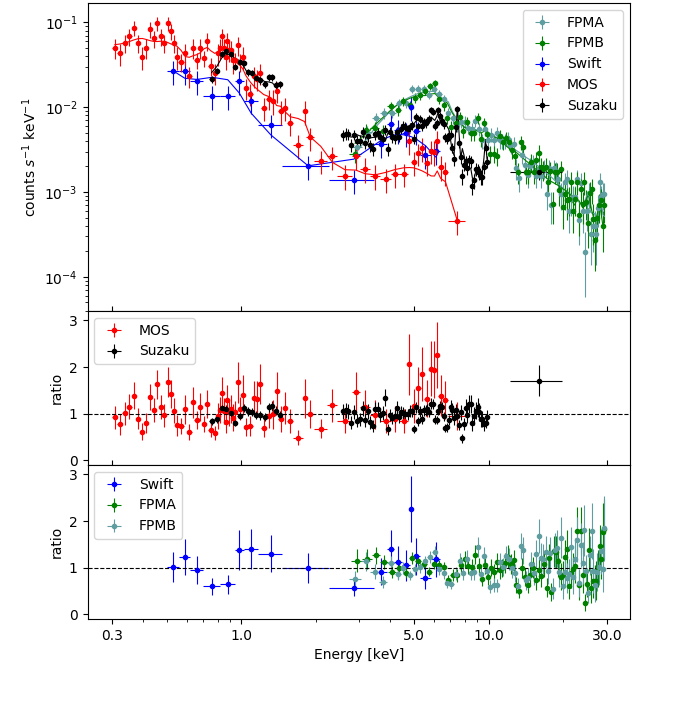}
	\end{minipage}
	\caption{\textit{Upper panel:} The theoretical best-fit ionised partial covering model shown over the entire X-ray band. The cyan line shows the continuum power law component with only Galactic absorption applied. \textit{Upper middle panel:} The folded spectra with the best fit partial cover model applied. \textit{Lower middle panel:} The ratio of data to model for the historic epoch, which includes the MOS, XIS and PIN spectra. \textit{Lower panel:} The ratio of data to model for the modern epoch, which contains the XRT and FPMA/FPMB spectra. }
	\label{fig:refcorr2}
\end{figure}

We adopt a partial covering scenario assuming the absorbers are in a compact region and only the primary continuum (power law) source is absorbed. We include two absorbers, one that is neutral and one with some non-zero ionisation parameter. This is not necessarily two distinct absorbers, but could be representative of a ionisation and density gradient within one medium. The {\sc xspec} model reads: {\sc constant$\times$tbabs(mekal + xillver + zxipcf$\times$zpcfabs$\times$powerlaw)}. 

Three scenarios were tested to explain the variability between epochs: (i) constant power law and varying absorbers; (ii) constant absorbers and vary power law; and  (iii) absorbers and power law free to vary. The continuum parameters were $\Gamma$ and the power law normalization. The absorber parameters were both covering fractions (CF), both column densities  ($\nh$) and the cloud ionization (${\rm log} \xi$). 

The initial model, with all parameters linked between epochs renders a very poor fit ($\cdof= 1563.29/603$), and the cross-calibration constants for FPMB is far too large to be acceptable ($\approx$1.7). Allowing the absorber parameters free to vary (Test (i)), we obtain a significantly better fit. If the CF is free at each epoch the fit improves by  $\dec= 860.28$ for two additional free parameters. If we allow the ${\rm log} \xi$ to be free between epochs as well the fit improves by $\dec= 39.97$ for two additional free parameter. Allowing all absorber parameters to be free between epochs gives the best fit, with $\cdof= 645.50/598$. 

Next, we test a variable continuum with a constant absorber (Test (ii)). Allowing just the power law normalization free between epochs (i.e. linked $\Gamma$), the fit quality is $\cdof=850.13/602$. Alternatively, permitting $\Gamma$ to be free between epochs and the normalization linked, the fit quality improves by $\dec=38$. If both the photon index and normalization are allowed free between epochs the $\cdof= 764.24/601$, an improvement of $\dec=47.89$ over the previous fit for one additional free parameter.  

If all absorber and continuum parameters are allowed to vary between epochs (Test (iii)), the best fit is $\cdof = 645.29/596$, comparable to Test (i), where the power law parameters are linked between epochs and  absorber parameters are free to vary. The data and residuals (separated by epoch) can be seen in Figure $\ref{fig:refcorr2}$.  The best fit parameters can be seen in Table $\ref{tab:PC}$.  The measured photon index is $\Gamma\sim 2.3$.  Compared to \cite{Grupe+2004}, who reported  $\Gamma\approx4$, our  value of $\Gamma$ is much more agreeable with other NLS1s.   This highlights the advantage of the high-energy sensitivity provided by \nustar.  The covering fractions of the absorbers in both epochs are very high, each approximately 90 per cent covering. This is  consistent with the high degree of optical polarization seen in \cite{Martin+1983}, and is consistent with the CF found in \cite{Grupe+2004} of 0.995. 

Figure \ref{fig:refcorr2} (top panel) shows the results for the best fitting model applied to only the FPMA data set for clarity. A number of interesting features are revealed. The intrinsic power law is shown in cyan, and the absorbed power law is shown in blue. This reveals that a large amount of absorption is required to reproduce the observed spectral curvature using this interpretation. This is sensible, as we measure covering fractions near 1. Given this high level of absorption at low energies, the {\sc mekal} component (shown in red) is clearly visible in the total model (shown in dark blue). In particular, the strong $0.9\kev$ feature found by \cite{Grupe+2004} is likely explained by this collisionally ionised plasma. This $0.9\kev$ feature is produced primarily by Fe L transitions. Contributions from the torus are shown in gray, and emission features at low energies also contribute to the soft excess. 

The second panel of Figure \ref{fig:refcorr2} shows the folded spectrum along with the models for each instrument. The colours are the same as those used in Figure \ref{fig:eeuf}. The model appears to provide a good fit to the data. This is more evident in the bottom panels, where the residuals (data/model) are shown. While some excess residuals are present in the $5-7\kev$ band in the MOS data, the model is clearly able to explain the overall shape of the spectra at each epoch. In particular, no clear excess residuals are seen at low energies, suggesting that the starburst model provides a satisfactory fit to the data.   Effectively, the long term variability in Mrk~1239 can be explained largely by changes in a partial covering medium and a relatively constant power law. The excess residuals seen in the MOS data could be due in part to a poor fit of a blurred \feka\ line suggesting a blurred reflection model must be examined. 
\def\arraystretch{0.95}
\begin{table*}
	\centering
	\begin{tabular}{c c c c c }
		\hline
		(1) 				 & (2) 				 & (3) 					& (4) 					&	(5)\\
		Model				 & Model Component	 & Model Parameter 		& \swift/ \nustar   	&	MOS/ \suzaku		\\
		\hline
		Ionized Partial Covering & Ionized absorber  & nH ($\num{e+22} {\rm cm^{-2}} $)	&  $64^{+37}_{-16}$	 	& $114^{+31}_{-56}$		\\
		& {\sc zxipcf} 	 &  ${\rm log} \xi$ 	[ergs cm  ${\rm s^{-1}}$]	& $2.8^{+0.1}_{-0.2}$	& $2.7^{+0.2}_{-0.4}$	\\ 
		& 					 & CF 					& $> 0.86$ 			 	& $0.9\pm 0.1$ 			\\ 
 \hhline{~----}
		& Neutral absorber  & nH ($\num{e+22} {\rm cm^{-2}} $)	& $13^{+10}_{-7}$ 	 	& $60^{+19}_{-17}$		\\ 
		& {\sc zpcfabs} 	 & CF 					& $0.96^{+0.03}_{-0.04}$& $0.89^{+0.04}_{-0.10}$\\ 
\hhline{~----}
		& 	Intrinsic 		 & $\Gamma$ 			& $2.3^{+0.3}_{-0.1}$	& $2.32 \pm 0.13$		\\ 
		& 	Power Law 		 & norm$^p$ ($\num{e-3}$)	& $6^{+4}_{-1}$ 		& $5^{+4}_{-3}$			\\ 
\hhline{~----}
		& FPMB	 Calibration & constant 			& $ 1.00 \pm 0.06$		&					\\
		& Swift Calibration & constant 			& $1.0 \pm 0.2$			&						\\ 
 \hhline{~----}
		& Collisionally Ionized material& kT [\kev]&  	-					&$0.66 \pm 0.03$		\\   
		& {\sc mekal}  	 & norm $(\num{e-4})$ [${\rm cm^{-5}}$]	& 	-					&	$1.3 \pm 0.1$ 		\\ 
		&    &Metal Abundance  [cosmic]   &       $1^f$  \\
\hhline{~----}
		& Distant  Cold Reflector   & norm ($\num{e-5}$)	& 	-					&$1.6 \pm 0.6$			\\ 
		& {\sc xillver}  	& ${\rm A_{Fe}}$ [solar]& $1^f$  \\
		& 					& ${\rm log} \xi$ [ergs cm  ${\rm s^{-1}}$]& $0^f$ \\
		& 					& ${\rm E_{cut}}$ [\kev] & $300^f$ \\
		&     				& Inclination $[^{\circ}]$		& $30^f$					\\ 	
\hhline{~----}%
\hhline{~----}
		Unabsorbed   Flux &(0.1-100 \kev)	 & $\num{e-11}$ [ergs ${\rm cm^{-2}}$  ${\rm s^{-1}}$]&5.74		&5.02\\
		Observed   Flux &(0.5-10 \kev)	 	 & $\num{e-11}$ [ergs ${\rm cm^{-2}}$  ${\rm s^{-1}}$]&0.289	&0.137\\
		Fit quality 		 & \cdof\ 			 &						&						&645.29/596 \\
		\hline
		\hline
		Blurred reflection		 & Neutral absorber  & nH ($\num{e+22} {\rm cm^{-2}} $)	&  $22 \pm 5$ 			& $61 \pm 11$\\ 
		& {\sc zpcfabs}     & CF 					& $0.99 \pm 0.01$		& $0.96^{+0.02}_{-0.01}$\\ 
		 \hhline{~----}
		& 	Intrinsic 		 & $\Gamma$ 			& $2.45^{+0.11}_{-0.14}$& $2.4 \pm 0.1$\\ 
		& 	Power Law 		 & norm ($\num{e-4}$)	&  $<22$				& $<18$\\ 
		 \hhline{~----}
		& FPMB Calibration   & constant 			& $1.00 \pm 0.06 $		& \\
		& Swift Calibration  & constant 			& $1.0 \pm 0.2$	    	&\\ 
		 \hhline{~----}	
		& Collisionally Ionized material& kT [\kev]& 		-				& $0.65 \pm 0.04$\\
		& {\sc mekal}  	& norm ($\num{e-5})$ [${\rm cm^{-5}}$]	& -						& $9 \pm 2$\\ 		
		&    &Metal Abundance  [cosmic] &       $1^f$  \\	
		 \hhline{~----}
		& Distant  Cold Reflector  & norm ($\num{e-5}$)	& -						& $2.1 \pm 0.06$ \\
		& {\sc xillver}  	& ${\rm A_{Fe}}$ [solar]& $1^f$  \\
		& 					& ${\rm log} \xi$ [ergs cm  ${\rm s^{-1}}$]& $0^f$ \\
		& 					& ${\rm E_{cut}}$ [\kev] & $300^f$ \\
		&     				& Inclination $[^{\circ}]$		& $<19^l$					\\ 	
		 \hhline{~----}
		& Blured reflector   &    ${\rm q_{in}}$      &        $3^f$      \\						
		&{\sc relxill} 		 &    ${\rm q_{out}}$      &        $3^f$      \\ 				
		&					 &   Break radius  [r$_{\rm g}$]       &       $6^f$      \\
		&					 &    spin           &       $0.998^f$  \\      
		&					 &  Outer Radius   [r$_{\rm g}$]       &        $400^f$    \\ 
		&					 & Inclination $[^{\circ}]$		& -						&  $<19$\\ 		
		&  	 & ${\rm log} \xi$	 [ergs cm  ${\rm s^{-1}}$]		&-						& $3.00^{+0.07}_{-0.25}$\\ 
		& 			  		 & ${\rm A_{Fe}}$ [solar]	& -		&$4^{+2}_{-1}$\\ 
		& 					& ${\rm E_{cut}}$ [\kev] & $300^f$ \\						
		& 				  	 & norm ($\num{e-5})$	& $10^{+6}_{-5}$		& $4 \pm 1$\\ 
		 \hhline{~----}
		 \hhline{~----}
		Unabsorbed	Flux & (0.1-100 \kev)	 	 & $\num{e-11}$ [ergs ${\rm cm^{-2}}$  ${\rm s^{-1}}$]&9.85 &4.32\\
		Observed   Flux  & (0.5-10 \kev)	 	 & $\num{e-11}$ [ergs ${\rm cm^{-2}}$  ${\rm s^{-1}}$]&0.289&0.134\\
		Fit quality  			 &\cdof				 & 						&						& 615.64/598\\
		\hline
	\end{tabular}
	\caption{Best-fit model parameters for Mrk~1239. Column (1) indicated the tested model and column (2) indicates the model component. Column (3) gives the value of each parameter for the modern epoch. Column (4) gives the value for each parameter for the historic epoch.  If a dash is present it indicated that the parameters are linked between epochs. All parameters with the superscript `$f$' are kept fixed at quoted values. The parameters with superscript `$l$' are linked between components. Normalizations with superscript `$p$'  are given by photons $\kev^{-1} {\rm cm}^{-2} s^{-1}$ at 1 keV.}
	\label{tab:PC}
\end{table*}
\subsubsection{Blurred Reflection}
\label{sec:refl}
 While there is a high degree of absorption in Mrk~1239, the variability in the $3-10\kev$ band suggests we may be probing emission from the inner black hole region.  Consequently, we examine if it is possible to model the intrinsic emission with a combination of power law and blurred reflection as has been done it some other highly absorbed, type II systems (e.g. \citealt{Walton+2019}).

In blurred reflection, photons emitted by the corona are then incident on the accretion disk where the are absorbed and re-emitted via florescence (e.g. \citealt{Ballantyne+2001}; \citealt{RossFabian+2005}). This reflected spectra is then subject to the general relativistic effects that are at play in accretion discs near black holes (e.g. \citealt{Miniutti+2004}). Blurred reflection has been successful in describing the spectral and timing properties  NLS1s (e.g. \citealt{Fabian+2009}; \citealt{Chiang+2015}; \citealt{GalloFabian+2012}, \citealt{Gallo+2015}; \citealt{Jiang+2019}; \citealt{Waddell+2019}). 

Here, the blurred reflection model, {\sc relxill} (\citealt{Garcia+2014}). replaces the ionized absorber in the partial covering scenario above. 
Given the data quality, the initial approach is rather conservative. The illumination as a function of distance (r) on the disc (emissivity profile) is described by a power law ($\propto {\rm r^{-q}}$) with index q.  For simplicity,  $q=3$, implying the primary emitter is radiating isotropically.  The cut off energy ({\rm $E_{cut}$}) is frozen at 300 \kev, as Mrk 1239 shows no evidence for a high energy cut off. The photon index is linked to the index of the primary power law component. The inner radius is fixed at the inner most stable orbit (ISCO) and the outer radius is fixed at 400r$_{\rm g}$. The dimensionless spin parameter defined by ${\rm a = cJ/GM^2}$ where M is the black hole mass and J is the angular momentum can take on values of 0 (non spinning) to 0.998 (maximum spin). The spin is linked between epochs and initially frozen at 0.998, given the complexities in measuring this parameter (e.g. \citealt{Bonson+Gallo+2016}). The inclination is linked with that of {\sc xillver}, which describes the cold distant reflector (i.e. torus).  It is linked between epochs, but left free to vary.  The disc ionization parameter (${\rm log \xi}$) and the iron abundance (${\rm A_{Fe}}$) are free to vary, but also linked between epochs. 

We adopt very similar geometry as the partial covering model, where the very central region of the AGN is highly obscured, there is some amount of cold distant reflection and a star forming region at some large distance from the AGN. The {\sc XSPEC} model used is {\sc constant$\times$tbabs$\times$(mekal + xillver + zpcfabs$\times$(relxill + powerlaw))}. As in the partial covering model we test three scenarios: (i) constant continuum and varying absorbers; (ii) constant absorbers and vary continuum; and (iii) continuum and absorber parameters free to vary. Here, continuum refers to the power law and blurred reflection components together.

\begin{figure}
	\begin{minipage}{0.48\textwidth}
		\includegraphics[width=\columnwidth]{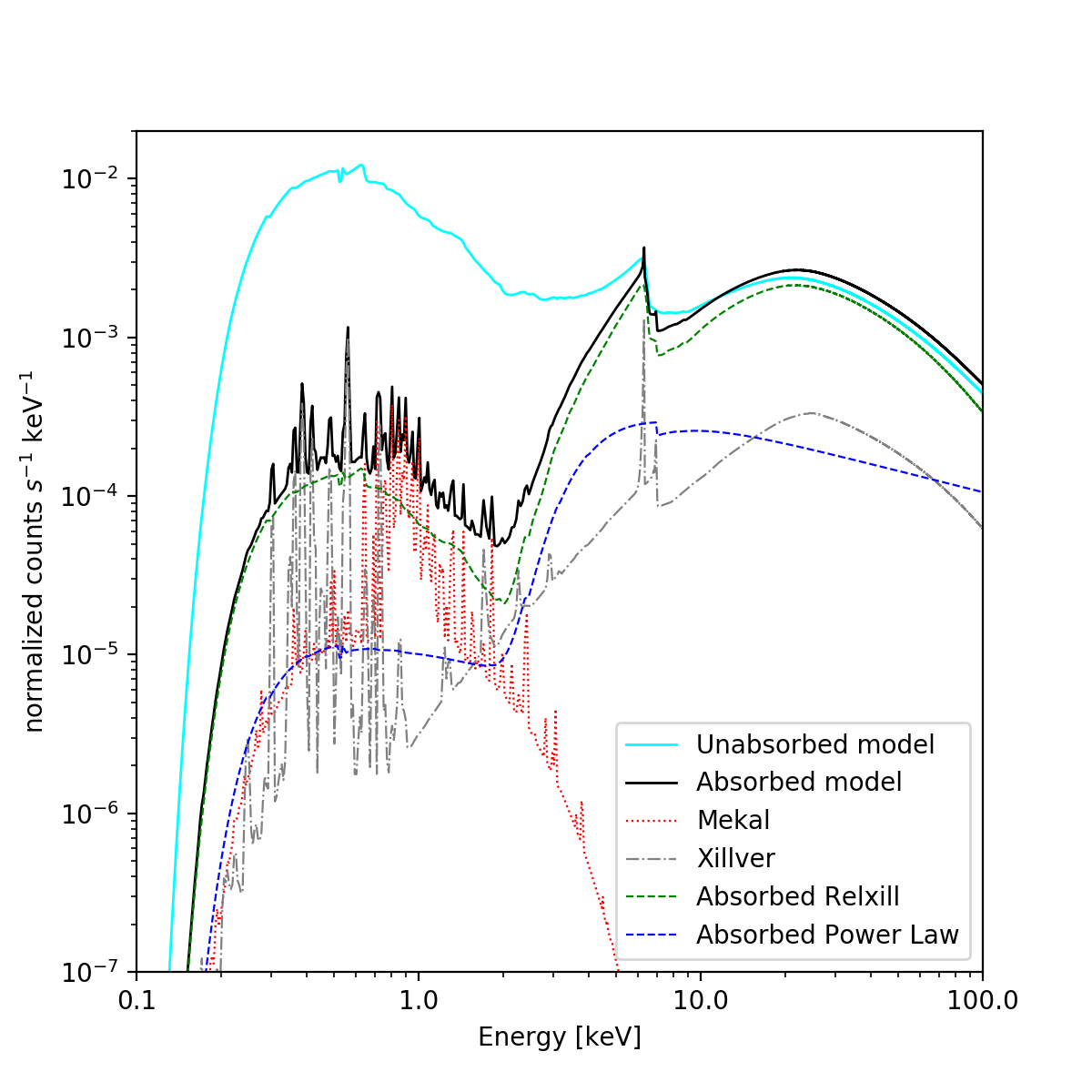}
	\end{minipage}
	\begin{minipage}{0.48\textwidth}
		\includegraphics[width=\columnwidth]{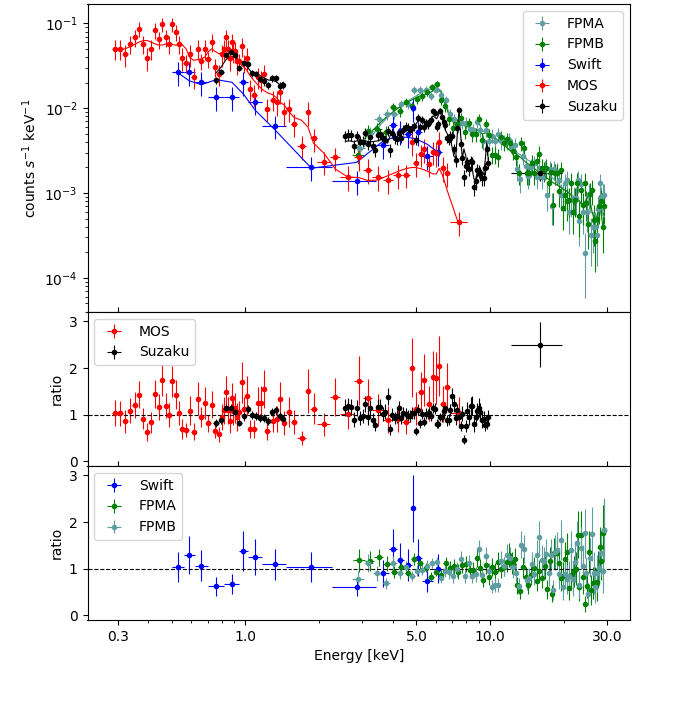}
	\end{minipage}
	\caption{\textit{Upper panel:} The theoretical best-fit  blurred reflection model shown over the entire X-ray band. The cyan line shows the continuum power law and blurred reflector components with only Galactic absorption applied. This model appears to be reflection dominated. \textit{Upper middle panel:} The folded spectra with the best fit blurred reflection model applied. \textit{Lower middle panel:} The ratio of data to model for the historic epoch, which includes the MOS, XIS and PIN spectra. \textit{Lower panel:} The ratio of data to model for the modern epoch, which contains the XRT and FPMA/FPMB spectra. }
	\label{fig:blur_RA}
\end{figure}

The initial model, with all parameters linked between epochs renders a very poor fit ($\cdof= 1538.17/603$).  Examining Test (i), if the CF was allowed to vary between epochs the the fit statistic improves by  $\dec =841.92$ for two additional free parameter. If instead $\nh$ is allowed to vary between epochs and CF is linked, the improvement to the fit is $\dec=45.2$. With both $\nh$ and CF free between epochs the $\cdof=645.42/599$.

In Test (ii), we linked absorbers and freed the continuum. If {\sc relxill} normalization was free, but linked between epochs, the fit statistic is $\cdof=720.31/602$. If instead the power law normalization was linked, but free to vary between epochs, there was no significant improvement in fit statistic. If both the {\sc relxill} and power law normalization are allowed free to vary between epochs the fit statistic is $\cdof=679.53/601$. 

Test (iii) renders the best fit  blurred reflection model.  The parameters are given in Table $\ref{tab:PC}$. The best fit blurred reflection model gave a fit statistic of $\cdof= 615.64/598$. This was a marked improvement over linking the continuum or absorption parameters separately. The reflection fraction at each epoch was calculated by determining the unabsorbed flux contributions between $0.1-100\kev$ in the blurred reflector ({\sc relxill}) and power law components.  Error on the calculated reflection fraction are propagated from the  uncertainties in the normalisation of each component.   Only a lower limit could be calculated at both epochs, as only an upper limit could be placed on the power law flux. The reflection fraction for the \nustar\ epoch is $>2.5$, for the MOS epoch $>1.3$, so the model is indicates the source is reflection dominated.  

Only an upper limit could be measured for the inclination ($<19^{\circ}$) indicating the system is nearly face-on.  \cite{Zhang+2002} measure an inclination of $7^{\circ}$ from the broadline region, which is in good agreement with our measurement.  

Figure $\ref{fig:blur_RA}$ (top panel) shows the theoretical best fit  blurred reflection model applied only to the FPMA data set for clarity.  The intrinsic continuum is shown in cyan, the absorbed power law and blurred reflection component are shown in blue and green, respectively. This again shows the large amount of absorption that is required for the {\sc mekal} (red) component to be visible. As in the ionized partial covering model, we believe that the {\sc mekal} is responsible for the 0.9 \kev\ feature reported by \cite{Grupe+2004}. The torus emission is  shown in gray and further contribute to the soft excess.  
The bottoms three panels of Figure $\ref{fig:blur_RA}$ show the  data and residuals (separated by epoch) for the best fit blurred reflection model, the colours are the same as Figures \ref{fig:eeuf} and \ref{fig:refcorr2}.  The fit has the same overall shape as the ionized partial cover model. We still see the residuals in the \feka\ band of the MOS spectra despite the fact that our model requires an over abundance of iron, which is commonly seen in NLS1. Over all the data fits well, and the soft band is again adequately  described by the {\sc mekal} component, and the long term variability is again driven by changes in the absorber. 

\section{Discussion}
\label{sec:Disc}
The data of Mrk 1239 analysed in this work highlight distinct physical process that dominate above and below $\sim3\kev$. These two regimes feature dramatically different variability properties and physical origins. As such our discussion shall be divided into these two energy regimes.  We interpret the spectra of Mrk 1239 to contain two distinct sources of emission, the first is the low energy star forming region and the second is the higher energy AGN component. 

\subsection{Origins of the soft excess}
To begin, our analysis of the Suzaku data did not give any evidence for variability below $3\kev$ on short timescales. As well the unfolded spectra of Mrk 1239 showed a remarkable consistency over the 18 years of spectra coverage presented in this paper, with virtually no change in flux in the 0.3-3 \kev\ while the harder band had significant change in flux over the same period. This is at odds with typical NLS1 where the soft band tends to have more variability (e.g. \citealt{Leighly1999}). As well, the spectral features present in the soft band, like the strong emission line-like features observed by \cite{Grupe+2004}, are inconsistent with smooth spectra typically seen in NLS1s. 

The long-term consistency of Mrk 1239 along with its unusual spectral shape lead us to apply different models to explain the soft excess. We use {\sc mekal} to model the possible contribution from starburst activity that may be present in Mrk 1239 (\citealt{Rod+2003}). This model has been used to describe many AGN that exhibit depressed power law emission from obscuration (e.g. \citealt{Franceschini+2003}) or intrinsic variability like Mrk~335 (e.g. \citealt{Gallo+2019m335}).  In AGN dominated flux states, this component is still present, but overwhelmed by the AGN emission.  

There is also a contribution to the line emission at low energies from the distant reflector (Figure \ref{fig:refcorr2} and \ref{fig:blur_RA}).  In combination, the {\sc mekal} and {\sc xillver} components nicely describe the line-like features in the spectra without requiring additional Gaussian profiles or abnormal abundances.  We considered the possibility that the emission could be due entirely to photoionized gas, perhaps from the narrow-line region. We tested this by replacing {\sc mekal} with a {\sc photemis} component to our best fit ionized partial covering model for just the MOS spectra between $0.3-2 \kev$. The model was only applied to the MOS spectra as it is computationally intensive. If there were any improvements over {\sc mekal}, they would be most obvious in the the MOS spectra. The best fit {\sc photemis} model resulted in a poorer fit than with {\sc mekal} ($\dec = 30$) for the same number of free parameter, and positive residuals in the $0.8-1 \kev$ band remained.

Table \ref{tab:flux_lumin} show the flux and luminosity for each component between $0.5-2\kev$ at each epoch. We can see that the {\sc mekal} component has a consistently strong contribution in both models with the central engine absorbed. But when the central engine absorption is removed, the line-emission {\sc mekal} is completely overwhelmed by the power law contribution in the partial covering scenario and by the power law and {\sc relxill} contribution in the blurred reflection scenario.

\def\arraystretch{1.5}
\begin{table*}
	\centering
	\begin{tabular}{c c c c c c c c }
		\hline
		(1) 		& (2) 		& (3) 		& (4) 		&	(5)		&	(6)		&	(7)		\\
		Model 		&Instrument	&Component	&Absorbed  Flux	&Absorbed  Luminosity	&Unabsorbed  Flux	&Unabsorbed Luminosity\\
		&			&			&[ergs ${\rm cm^{-2}}$  ${\rm s^{-1}}$]& [ergs   ${\rm s^{-1}}$]&[ergs ${\rm cm^{-2}}$  ${\rm s^{-1}}$]& [ergs ${\rm s^{-1}}$]\\
		\hline
		Blurred Reflection	&\swift/\nustar&Power Law&$\num{2.20E-14}$&$\num{1.98E+40}$&$\num{1.88E-12}$&$\num{1.71E+42}$	\\
		&			&Relxill	&$\num{17.0E-14}$	&$\num{15.6E+40}$	&$\num{1.52E-11}$	&$\num{1.40E+43}$	\\
		&			&Xillver	&$\num{7.39E-14}$	&$\num{6.73E+40}$	&$\num{9.36E-14}$	&$\num{8.55E+40}$	\\
		&			&Mekal		&$\num{11.6E-14}$	&$\num{10.5E+40}$	&$\num{1.30E-13}$	&$\num{1.17E+41}$	\\
		 \hhline{~------}
		&MOS/\suzaku	&Power Law	&$\num{6.13E-14}$	&$\num{5.53E+40}$	&$\num{1.85E-12}$	&$\num{1.67E+42}$	\\
		&			&Relxill	&$\num{19.6E-14}$	&$\num{18.0E+40}$	&$\num{6.13E-12}$	&$\num{5.64E+42}$	\\
		&			&Xillver	&$\num{7.18E-14}$	&$\num{6.48E+40}$	&$\num{9.10E-14}$	&$\num{8.23E+40}$	\\
		&			&Mekal		&$\num{11.6E-14}$	&$\num{10.5E+40}$	&$\num{1.30E-13}$	&$\num{1.18E+41}$	\\
		\hline
		\hline
		Partial Covering&\swift \nustar&Power Law&$\num{16.1E-14}$&$\num{14.9E+40}$	&$\num{1.38E-11}$	&$\num{1.25E+43}$	\\
		&			&Xillver	&$\num{4.97E-14}$	&$\num{45.1E+40}$	&$\num{6.25E-14}$	&$\num{5.70E+40}$	\\
		&			&Mekal		&$\num{17.0E-14}$	&$\num{15.3E+40}$	&$\num{1.90E-13}$	&$\num{1.71E+41}$	\\
		 \hhline{~------}
		&MOS/ \suzaku	&Power Law 	&$\num{22.5E-14}$	&$\num{20.6E+40}$	&$\num{1.21E-11}$	&$\num{1.09E+43}$	\\
		&			&Xillver	&$\num{4.86E-14}$	&$\num{43.8E+40}$	&$\num{6.12E-14}$	&$\num{5.53E+40}$	\\
		&			&Mekal		&$\num{17.0E-14}$	&$\num{15.3E+40}$	&$\num{1.90E-13}$	&$\num{1.72E+41}$	\\
		\hline
	\end{tabular}
	\caption{Flux and luminosity for each model component between $0.5 -2 \kev$. Column (1) indicates the model used for the measurements. Columns (2) and (3) denotes the instrument used/the epoch and the model component measured respectively. Columns (4) and (5) list the absorbed flux and luminosity between $0.5 -2 \kev$. Columns (6) and (7) list the unabsorbed flux and luminosity between $0.5 -2 \kev$. There are small discrepancies in the flux and luminosity measured between epochs in the same model even though the xillver component are shared between epochs. This is due to the differences in telescope responses.}
	\label{tab:flux_lumin}
\end{table*}

We are using {\sc mekal} to describe the X-ray evidence of a starburst activity. Mrk 1239  has significant evidence of star burst activity. The PAH signatures found by \cite{Rod+2003}  agrees with this interpretation. The starburst regions are included in the X-ray extraction regions of the instruments used due to the modest angular resolution. 

A study of ten ultraluminous infrared galaxies (ULIRGs) by \cite{Franceschini+2003} used {\sc mekal} to model the SFR in  selected galaxies. They measured the average temperature of the {\sc mekal} component to be ${\rm kT}\approx 0.7 \kev$ which agrees exactly with the temperature measured in Mrk 1239 (${\rm kT}\approx 0.66 \kev$). The luminosity we measure for the {\sc mekal} component also agrees with the luminosity \cite{Franceschini+2003} measured in their sample. 

\cite{Franceschini+2003} also give a method to approximate the SFR using $L_{2-10 \kev}$:
\begin{equation}
{\rm SFR^{ULIRG}_{X-ray}} \approx \frac{L_{2-10 \kev}}{10^{39} {\rm erg\; s^{-1}}} \Msun {\rm yr^{-1}}
\end{equation}
If we measure the $L_{2-10 \kev}$ for just the unabsorbed {\sc mekal} component we get a SFR of 5.8 and 3.7 $ \Msun {\rm yr^{-1}}$ based on the ionized partial cover and blurred reflection model respectively.  This agrees nicely with the SFR predicted by the PAH signatures (\citealt{Ruschel_Dutr}).

Another interesting feature of {\sc mekal} component is that it appears to be  independent of the continuum model tested. The {\sc mekal} temperature and normalization agree with each other to within 1 and $2\sigma$, respectively. This suggests that regardless of the mechanism producing the observed curvature at high energies, the soft X-ray emission of Mrk 1239 can be nicely explained by the starburst component and some photoionised emission from distant emission, without the need for significant overabundances of Ne or other elements.  We found that the spectral feature \cite{Grupe+2004} reported as a Ne overabundance was a blend of many Fe L transitions.

We found that the spectral feature \cite{Grupe+2004} reported as a Ne overabundance was a blend of many Fe L transitions. We examined if adopting different cosmic abundances in {\sc XSPEC} could alter the fit quality. Regardless, of the abundance table used, the {\sc MEKAL} temperature remained between $0.63-0.65 \kev$ and the fit quality was ${\rm C} = 162-167$ for 165 dof.
\subsection{The hard X-ray spectrum of Mrk 1239}
The light curves seen in Figure \ref{fig:refcorr} show rapid variability between $3-10\kev$. Similarly, the multi-epoch spectra in Figure \ref{fig:eeuf} suggest long term variability between $3-10\kev$. 
In Section~\ref{sect:var}, the excess variance in the \nustar\ light curve was calculated and determined to be $\sigma^{2}_{\rm rms}=0.04\pm0.01$.  With the caveats that our analysis uses slightly different energy bands and time bins, the measured value of $\sigma^{2}_{\rm rms}$ for Mrk~1239 is comparable to the average value \cite{Ponti+2012} measured for NLS1s ($\sigma^{2}_{\rm NLS1} \approx 0.02$).  Following the ${\rm M_{BH}}$-$\sigma_{\rm rms}$ relation found by \cite{Ponti+2012}, we estimate a black hole mass  of $2\times10^6$ \Msun\ for Mrk~1239, which is in agreement with other works (see \citealt{Ryan+2007}).  Based on the rapid variability in the $3-10\kev$ band, Mrk~1239 behaves like an unobscured NLS1.  The central black hole region is exposed in the $3-10\kev$ band.

Independent of the scenarios tested (i.e. partial covering or blurred reflection), the photon index of the intrinsic power law component is comparable ($\Gamma\sim2.3$).  The measured value is much flatter than has been measured in other works (\citealt{Grupe+2004}), but it is more consistent with what is expected from Seyfert galaxies and NLS1s (\citealt{Grupe+2010}).  Estimating the Eddington luminosity ratio from $\Gamma$ based on the relationship in \cite{Brightman+2013} gives L/L$_{Edd} \sim1-1.5$ for Mrk~1239. This is consistent with the value estimated by \cite{Yao+2018}. The high value implies a rapid accretor, which is also characteristic of the  NLS1 class.  Alternatively, the value could be overestimated if the primary emission is anisotropic, which might be consistent with the jet interpretation for some of the radio emission (\citealt{Do+2015}).

Comparing the partial covering and blurred reflection scenarios directly, similar neutral absorbers that are responsible for obscuring the low energy X-rays and revealing the starburst regions are required in both models.  In both models the column density decreases between the historic and modern epochs.  The main difference in the two scenarios is the strength of the primary power law.   In the blurred reflection scenario the spectra is extremely reflection dominated with the reflection fraction $>1.3$ and $>2.5$ for the historic and modern epoch, respectively.  The reflection dominated interpretation may be at odds with the value of the emissivity index being fixed to ${\rm q_{in}}=3$ (allowing it to be free did not improve the fit), which would imply an inner disk that is truncated at a large radius or a corona that is very high above the disc.  A high reflection fraction requires a significant amount of light bending or a non-standard geometry for the accretion disc. 

Though it appears clear that we are observing emission originating close to the black hole, it is difficult to produce a self-consistent blurred reflection model with the current data.  The ionized partial covering scenario has the advantage of being consistent with the high levels of polarization seen in the optical band by \cite{Martin+1983} and \cite{Goodrich+1989}.

In both of the scenarios, the primary emission variability is minimal on long timescales. The change in flux above 3 \kev\ observed on long timescales is driven by the absorbers. A possible explanation for the large amount of absorbing material surrounding the AGN could be that it contains a Schwarzschild black hole. According to \cite{Ishibash+2020}, non-spinning black holes have less radiation pressure to drive obscuring gas and dust out of our line of sight than Kerr black holes do. The best fit blurred reflection model with $a=0$ results in a fit of $\cdof=625.50/598$. Most fit parameters remained  similar to the values we report in Table~\ref{tab:PC}, except for the iron abundance which decreases to ${\rm A_{Fe}}=2.2$. Although the fit is not as good as that produced by the Kerr black hole model, a non-spinning black hole is more consistent with the low emissivity profile that is adopted.

\begin{figure*}
	\begin{center}
		\includegraphics[width=1.8\columnwidth]{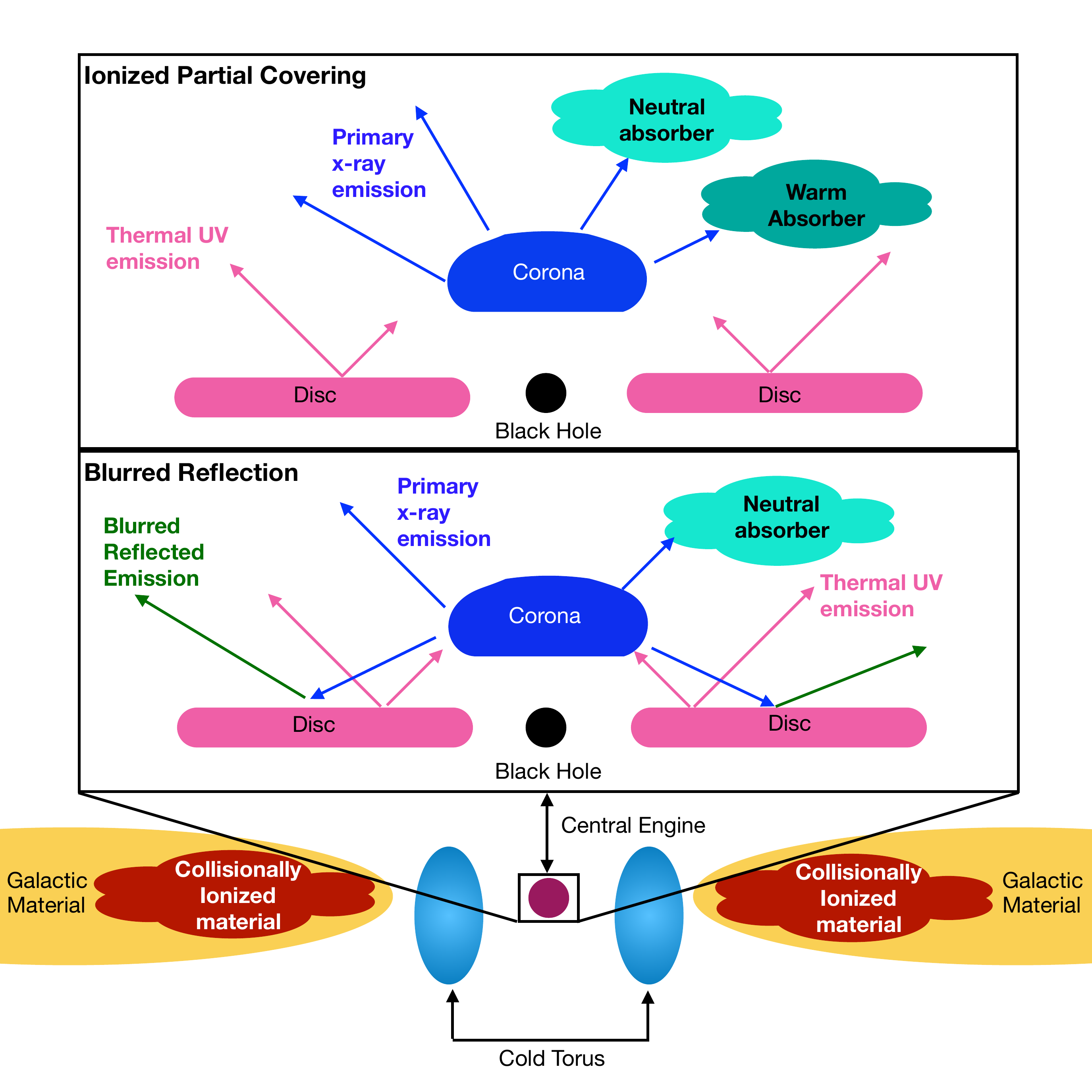}
		\caption{A graphic interpretation of our proposed models. The top box shows the AGN in the ionized partial covering scenario. The bottom box show the AGN in the blurred reflection scenario. In both the ionized partial covering and blurred reflection scenarios, the absorption of the primary continuum is inside the torus. }
	\label{fig:Dia2020}
\end{center}
\end{figure*}

Figure \ref{fig:Dia2020} show a diagram for the physical scenario we propose. The top box shows the physical scenario for the ionized partial covering model. Here photons are emitted by the disc in the UV and are Compton-upscatted by the corona. The primary X-ray photons are then emitted by the corona, where they encounter the partial covering components. This includes a neutral partial cover and an ionized partial cover (warm absorber). In our model most of the photons interact with the absorbers, and few escape the inner region unaffected. This is due to the high CF we measure. The bottom of the two boxes show the AGN for the blurred reflection scenario. Here UV photons are produced by the disk, Compton-upscattered by the corona and then reflected off of the disc. Both the primary and blurred reflected emission are absorbed by a neutral absorber before leaving the AGN. Once we move out to the torus our diagrams for the ionized partial cover and blurred reflection scenario are the same. The photons from the AGN interact with the torus to give us cold distant reflection emission lines. Then our collisionally ionized material, which we believe is starburst activity and therefore must be located within 400 pc of the central engine as PAH measurements show starburst activity here (\citealt{Ruschel_Dutr}). Outside of the starburst region we see no evidence of further emission in the X-ray band. Alternatively the cold absorber could be associated with the torus.  Perhaps our line-of-sight is grazing the torus or the torus is more patchy and isotropic around the primary source.

The difference in each of these scenarios is how the primary continuum is produced and then absorbed. If the absorbing material is in fact surrounding the AGN then it stands to reason that the absorbing material is what is driving the long term variability in Mrk 1239. This is the case regardless of which model (ionised partial covering or blurred reflection) is implied as the additional component to fit the continuum above $3\kev$.
\section{Conclusion}
\label{sect:conclusion}
In this work, multi-epoch X-ray spectra spanning 18 years of the NLS1 galaxy Mrk 1239 are presented. This study combines data from \xmm\ and \suzaku\ with a simultaneous \nustar\ and \swift\ observation to study the spectrum and variability of this source. A comparison of the unfolded spectra reveals that the spectra are very similar at all epochs below $3\kev$, while the \nustar\ and \swift\ data are brighter by a factor of $2-3$ at higher energies. The light curves also reveal no short term variability below $3\kev$, while modest variability at higher energies is seen. 

Motivated by this, and by signatures of star formation in the infrared spectra presented in previous works, we successfully model the soft spectrum with emission from a hot plasma. When the absorption component is removed, the starburst component accounts for $\sim1-10$ per cent of the total emission, which is sensible for a type-1 AGN.
Both a blurred reflection and ionised partial covering model are employed to explain the remainder of the emission. Both models have several key components in common; a power law component with a slope of $\sim2.3-2.4$, a neutral absorber with a covering fraction near $\sim1$, and contributions from a distant reflector (e.g. the torus). There are some apparent inconsistencies which may be difficult to explain in the blurred reflection model, including a high iron abundance, and a low emissivity index despite having a reflection dominated spectrum. By contrast, the two-component absorption model may agree with the two polarisation regions found in the optical emission.

Deep observations with calorimeter-type resolution (e.g. \citealt{Takahashi+2016}) such as \xrism\ (\citealt{Tashiro+2018}) may help to reveal any emission line features present in the low energy spectrum from the starburst and distant reflection components.  They may also reveal absorption features present at high energies, helping to confirm the high level of absorption required to fit the observed curvature. Additionally, future observations with \xmm\ may help to confirm the minimal level of low energy variability and support a starburst emission component to explain the soft emission from this unique source.

\section*{Acknowledgements}
Many thanks to Adam Gonzalez for insightful discussions. The authors thank the anonymous referee for their comments.   The \xmm\ project is an ESA Science Mission with instruments and contributions directly funded by ESA Member States and the USA (NASA). This research has made use of data obtained from the \suzaku\ satellite, a collaborative mission between the space agencies of Japan (JAXA) and the USA (NASA). Additionally, this work made use of data supplied by the UK Swift Science Data Centre at the University of Leicester, data from the NuSTAR mission, a project led by the California Institute of Technology, managed by the Jet Propulsion Laboratory, and funded by the National Aeronautics and Space Administration. MZB, SGHW and LCG acknowledge the support of the Natural Sciences and Engineering Research Council of Canada (NSERC). LCG acknowledges financial support from the Canadian Space Agency (CSA).


\bibliography{bibfile}
\bibliographystyle{aasjournal}
\end{document}